\documentclass{iau}

\usepackage{amsmath}
\usepackage{graphicx}
\usepackage{multirow}
\usepackage{subcaption}

\usepackage{booktabs}

\let\lesssim\relax
\let\gtrsim\relax
\usepackage{amssymb}

\begin{document}

\lefttitle{Abbas Askar et al.}
\righttitle{Proceedings of the International Astronomical Union: No.398 and MODEST-25}

\jnlPage{1}{7}
\jnlDoiYr{2021}
\doival{10.1017/xxxxx}

\aopheadtitle{Proceedings IAU Symposium}
\editors{eds. Hyung Mok Lee, Rainer Spurzem and Jongsuk Hong}

\title{Formation and Growth of IMBHs in Dense Star Clusters: Lessons from \textit{N}-body and MOCCA Monte Carlo Simulations}

\author{Abbas Askar$^1$, Marcelo~C.~Vergara$^2$ and Sohaib Ali$^{1,3}$}
\affiliation{$^1$Nicolaus Copernicus Astronomical Center, Polish Academy of Sciences, Warsaw, Poland
\email{askar@camk.edu.pl}}
\affiliation{$^2$Astronomisches Rechen-Institut, Zentrum für Astronomie, University of Heidelberg, Mönchhofstrasse 12-14, 69120, Heidelberg, Germany}
\affiliation{$^3$Institute of Astronomy, Nicolaus Copernicus University, Jurija Gagarina 11, 87-100 Toruń, Poland}

\begin{abstract}
Dense star clusters are promising nurseries for the formation and growth of intermediate-mass black holes (IMBHs; $\rm\sim10^2$–$\rm 10^5\,\rm M_\odot$), with increasing observational evidence pointing to their presence in massive clusters and stripped dwarf-galaxy nuclei. During the early evolution of compact clusters, massive stars can rapidly segregate to the center, where frequent collisions may trigger the runaway growth of a very massive star (VMS). This object can subsequently collapse to form an IMBH or merge with a stellar-mass black hole. We use direct $N$-body and Monte Carlo simulations of clusters with initial core densities between $10^6$ and $4\times10^8\,\rm M_\odot\,\mathrm{pc}^{-3}$ and total masses of $\rm 5.9\times10^5$ and $\rm1.3\times10^6\,M_\odot$. These models show that IMBHs of $\rm10^3$–$\rm10^4\,M_\odot$ can form within $\lesssim5$ Myr through the runaway collision channel. At later times, the IMBHs continue to grow through mergers with black holes, stars, and compact remnants, providing predictions testable with future gravitational-wave and transient surveys.
\end{abstract}


\maketitle

\section{Introduction}\label{intro}

Intermediate-mass black holes (IMBHs; $\sim10^2$–$\rm 10^5\,M_\odot$) occupy the gap between stellar-mass BHs (sBHs; $\rm \lesssim100\,M_\odot$) and supermassive BHs (SMBHs; $\rm \gtrsim10^6\,M_\odot$). Observational evidence for their existence is mounting. In particular, fast-moving stars in the core of Omega Centauri are consistent with the presence of an IMBH \citep{Haeberle24}, while additional candidates have been found in dwarf galaxies and extragalactic clusters \citep{Lin18,Pechetti22}. The first confirmed detections of low-mass IMBHs ($\rm \sim10^2\,M_\odot$) have come from binary BH mergers observed in gravitational waves (GWs) by the LIGO–Virgo–KAGRA (LVK) collaboration, including GW190521 and GW231123 \citep{LVK25-gwtc4}.

Dense clusters, with short relaxation times and frequent stellar encounters, are favorable sites for seeding and growing IMBHs. Several pathways have been proposed \cite[see][for a detailed review and references therein]{AskarBaldassareMezcua24}, including runaway stellar collisions that form a very massive star (VMS) which may later collapse or merge with a sBH \citep[e.g.,][]{PZwart04,Freitag2006,Giersz15,dicarlo2021,fuji2024,rantala2024,Sharma2025,vergara2025}. Here we briefly summarize results from complementary $N$-body and MOCCA Monte Carlo simulations of compact, centrally concentrated star clusters. We examine the formation of a VMS through collisional runaway, the conditions for IMBH seeding, and its subsequent growth through stellar and compact-object mergers, while highlighting caveats and implications for gravitational-wave and high-energy observations.

\section{First study: Million-body compact cluster model}\label{sec:nbody}

We model the evolution of an extremely dense, isolated cluster with $N=10^6$ single stars and no primordial binaries at metallicity $Z=0.01$, using \textsc{Nbody6++GPU} \citep{vergara2025}. The cluster has $\rm M = 5.87\times10^5\,M_\odot$, a \citet{king1966} profile with $\rm W_0=6$, and half-mass radius $\rm r_h=0.1$ pc. Stars are initialized on the zero-age main sequence with masses between $0.08$ and $\rm 150\,M_\odot$, drawn from a two-component \citet{kroupa2001} initial mass function (IMF). For comparison, we also ran a MOCCA simulation with the same initial model\footnote{Data: \url{https://doi.org/10.5281/zenodo.15283075}}. These conditions yield a core density $\rm \rho_c=4\times10^8\,M_\odot\,\mathrm{pc}^{-3}$ and half-mass relaxation time of 8.6 Myr. Massive stars segregate on much shorter timescales, driving high encounter rates that lead to runaway collisions and formation of a VMS of $\rm \gtrsim5\times10^4\,M_\odot$ within $\lesssim5$ Myr.

\begin{figure*}[htbp]
  \centering
  \begin{subfigure}[t]{0.49\linewidth}
    \centering
    \includegraphics[width=\linewidth]{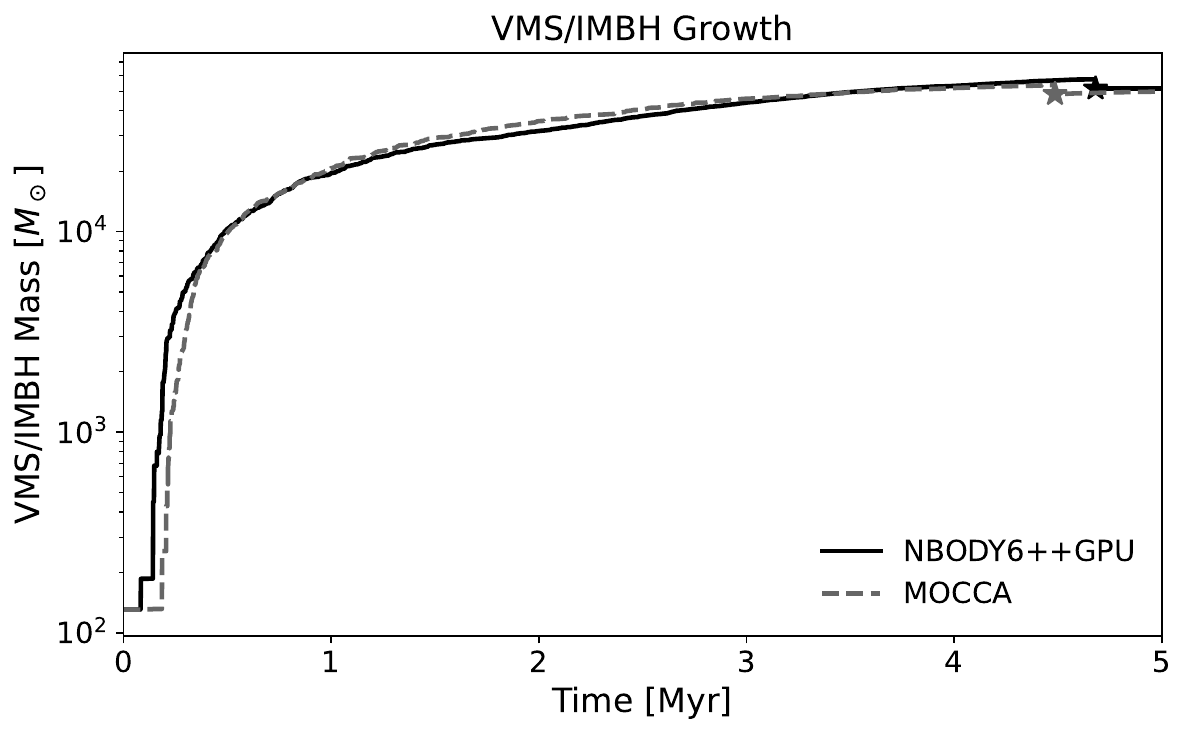}
    \caption{VMS/IMBH mass growth up to 5 Myr from \textsc{Nbody6++GPU} (solid) and MOCCA (dashed). The star marks IMBH formation.}
    \label{fig:vms_growth_bw}
  \end{subfigure}\hfill
  \begin{subfigure}[t]{0.49\linewidth}
    \centering
    \includegraphics[width=\linewidth]{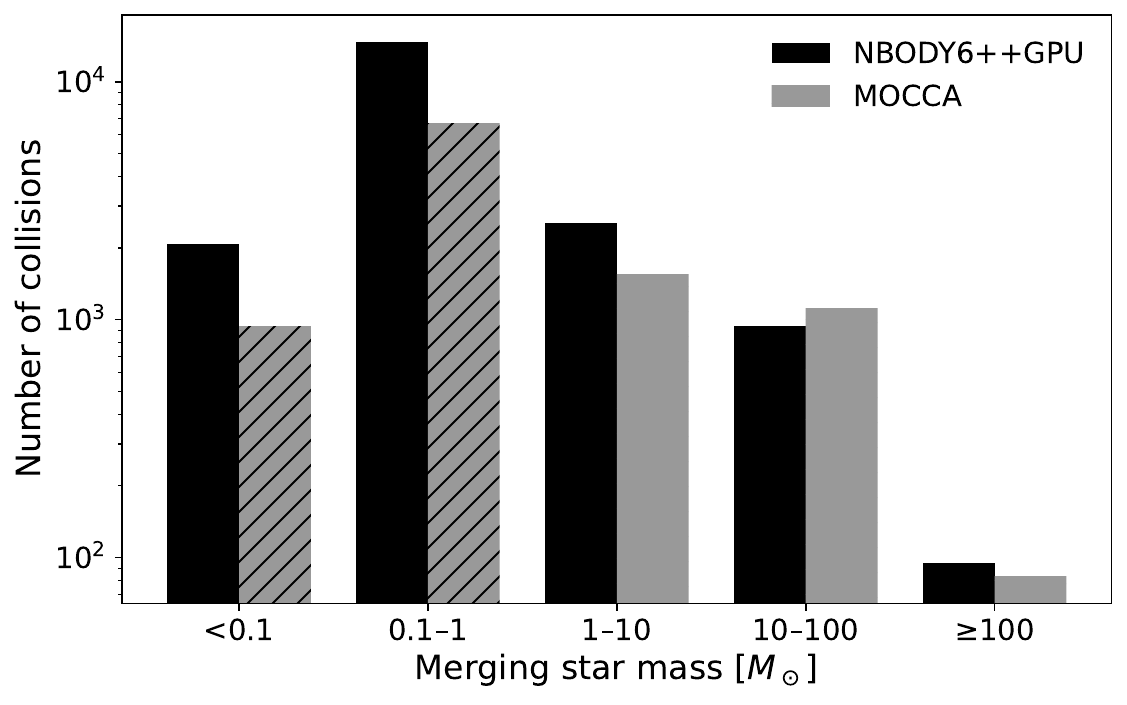}
    \caption{Mass distribution of stars colliding with the VMS during runaway growth, from \textsc{Nbody6++GPU} (black) and MOCCA (gray).}
    \label{fig:vms_collision_hist}
  \end{subfigure}
  \label{fig:vms_side_by_side}
\end{figure*}

Both simulations were evolved to 5 Myr. By $t\simeq4.5$ Myr, the runaway produced a VMS that collapsed into an IMBH at the cluster center in both runs. The VMS grew to $\rm \gtrsim5\times10^4\,M_\odot$ through successive collisions, most within the first 2–3 Myr (Fig.~\ref{fig:vms_growth_bw}). Stellar evolution followed updated SSE/BSE prescriptions \citep{hurley2002,kamlah2022}, with improved treatment of VMS properties, rejuvenation, and mass loss \citep{vergara2025}.

The two approaches agree on the overall growth pathway and timing of IMBH formation, but the $N$-body run produced nearly twice as many VMS collisions, mainly with low-mass stars. At higher masses, the distributions agree more closely, with both methods yielding on the order of $10^3$ collisions involving stars above $\rm 10\,M_\odot$ (Fig.~\ref{fig:vms_collision_hist}). These differences arise from MOCCA’s probabilistic treatment of collisions, which tends to underestimate the frequency of low-mass mergers in this type of cluster. As a result, IMBH formation occurs slightly earlier ($\sim4.5$ Myr vs.\ $\sim4.7$ Myr in the $N$-body run), owing to weaker rejuvenation of the VMS. As a cautionary point, collisions in the rapid early growth phase (first 2 Myr) occur on $200$–$1000$ yr timescales, shorter than the thermal timescale of the VMS ($\sim10^4$ yr). Consequently, the structure of the runaway product before collapse remains uncertain, and its inferred mass should be considered an upper limit.

\section{Second study: MOCCA simulations of dense clusters}\label{sec:mocca}

We also ran a suite of MOCCA simulations of dense clusters with $N=2\times10^6$ objects (10\% binaries), metallicity $Z=0.0005$, total mass $\rm M=1.29\times10^6\,M_\odot$, and circular orbits at 10 kpc. We adopted \citet{king1966} models with $\rm W_0\in\{5,8\}$ and varied $\rm r_h$ to span $\rm \rho_c=8.7\times10^5$–$\rm 4.4\times10^8\,M_\odot\,\mathrm{pc}^{-3}$ (Table~\ref{tab:mocca_models}). Stellar and binary physics included rapid SN/fallback \citep{fryer2012}, metallicity-dependent winds \citep{belczynski2010}, and updated low-$Z$ massive-star evolution \citep{Tanikawa20}. In BH–star collisions, 25\% of stellar mass was accreted; MS–MS collisions followed BSE rejuvenation; and GW recoil kicks were applied with natal BH spins sampled uniformly in $[0,0.1]$.

\begin{table}[htbp]
\centering
\caption{Key initial conditions and outcomes of the MOCCA models. All runs have $N=2\times10^6$ objects (10\% binaries), $Z=0.0005$, $\rm M=1.29\times10^6\,M_\odot$, and stars initialized on the ZAMS with masses $0.08$–$150\,M_\odot$ from a two-component \citet{kroupa2001} IMF. Here $\rm W_0$ is the King concentration parameter and $\rm T_{rh}$ is the initial half-mass relaxation time.}
\label{tab:mocca_models}
\begin{tabular}{lccccc}
 \toprule
 Model & $\rm r_h$ (pc) & $\rm W_0$ & $\rm \rho_{\rm core}$ ($\rm M_\odot\,\mathrm{pc}^{-3}$) & $T_{rh}$ (Myr) & IMBH? ($\rm M_{\rm BH}$ at 12 Gyr) \\
 \midrule
 rh0.25--W08 & 0.25 & 8 & $4.38\times10^8$ & 42.7 & \checkmark ($2.762\times10^4\,M_\odot$) \\
 rh0.5--W08 & 0.50 & 8 & $5.86\times10^7$ & 121.0 & \checkmark ($1.537\times10^4\,M_\odot$) \\
 rh0.5--W05 & 0.50 & 5 & $4.34\times10^6$ & 121.0 & $\times$ \\
 rh1.0--W08 & 1.00 & 8 & $7.36\times10^6$ & 342.3 & \checkmark ($6.244\times10^3\,M_\odot$) \\
 rh2.0--W08 & 2.00 & 8 & $8.67\times10^5$ & 965.6 & $\times$ \\
 \bottomrule
\end{tabular}
\end{table}

\begin{figure*}[htbp]
    \centering
    \begin{subfigure}{0.49\textwidth}
        \centering
        \includegraphics[width=\linewidth]{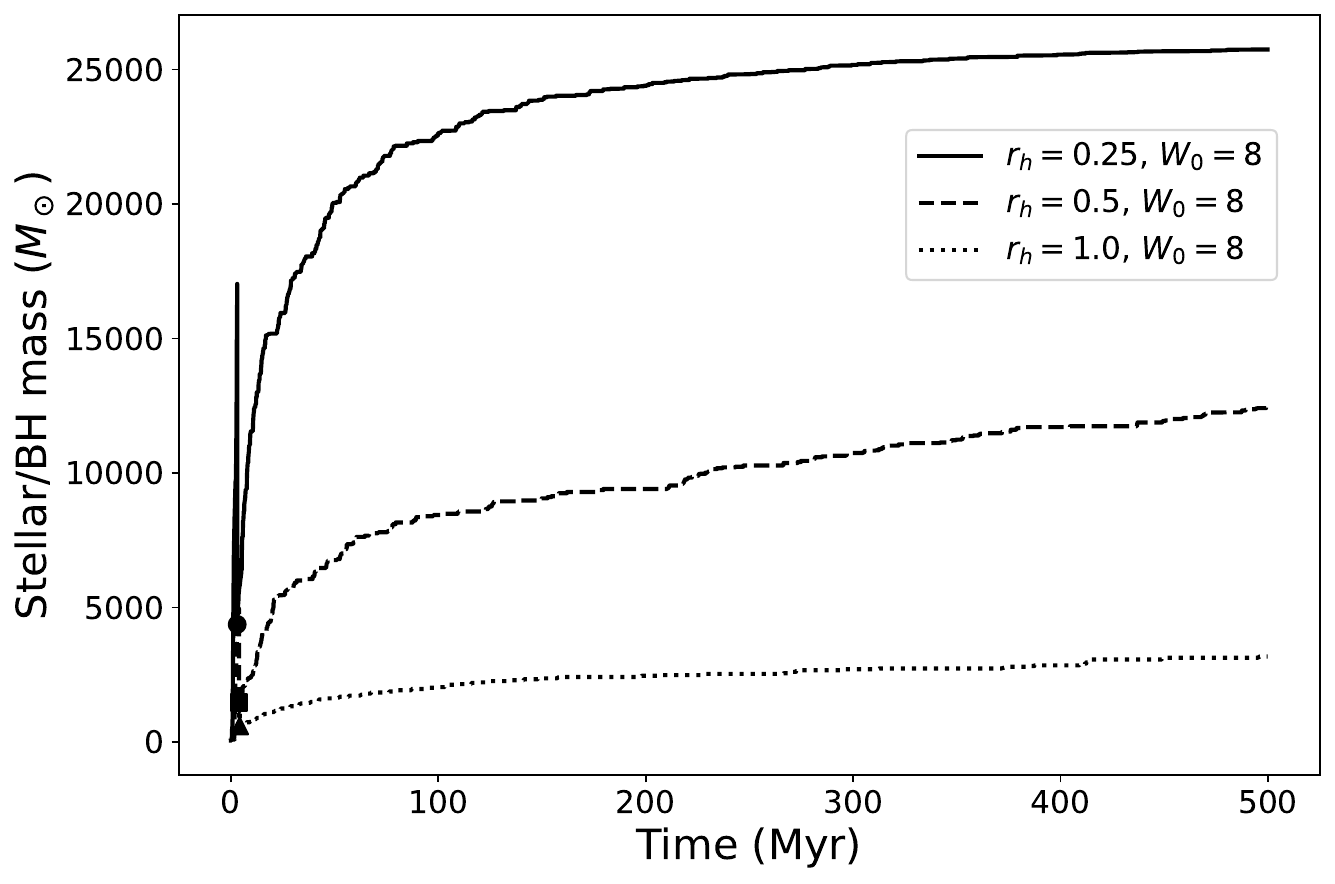}
       \caption{Early IMBH growth (first 500 Myr) in models with $\rm r_h=0.25$, 0.5, and 1.0 pc ($\rm W_0=8$). Symbols denote IMBH seeding via VMS--sBH mergers, with 25\% of the VMS mass accreted by the sBH.}
        \label{fig:imbh_mass_500Myr}
    \end{subfigure}
    \hfill
    \begin{subfigure}{0.49\textwidth}
        \centering
        {\includegraphics[width=\linewidth]{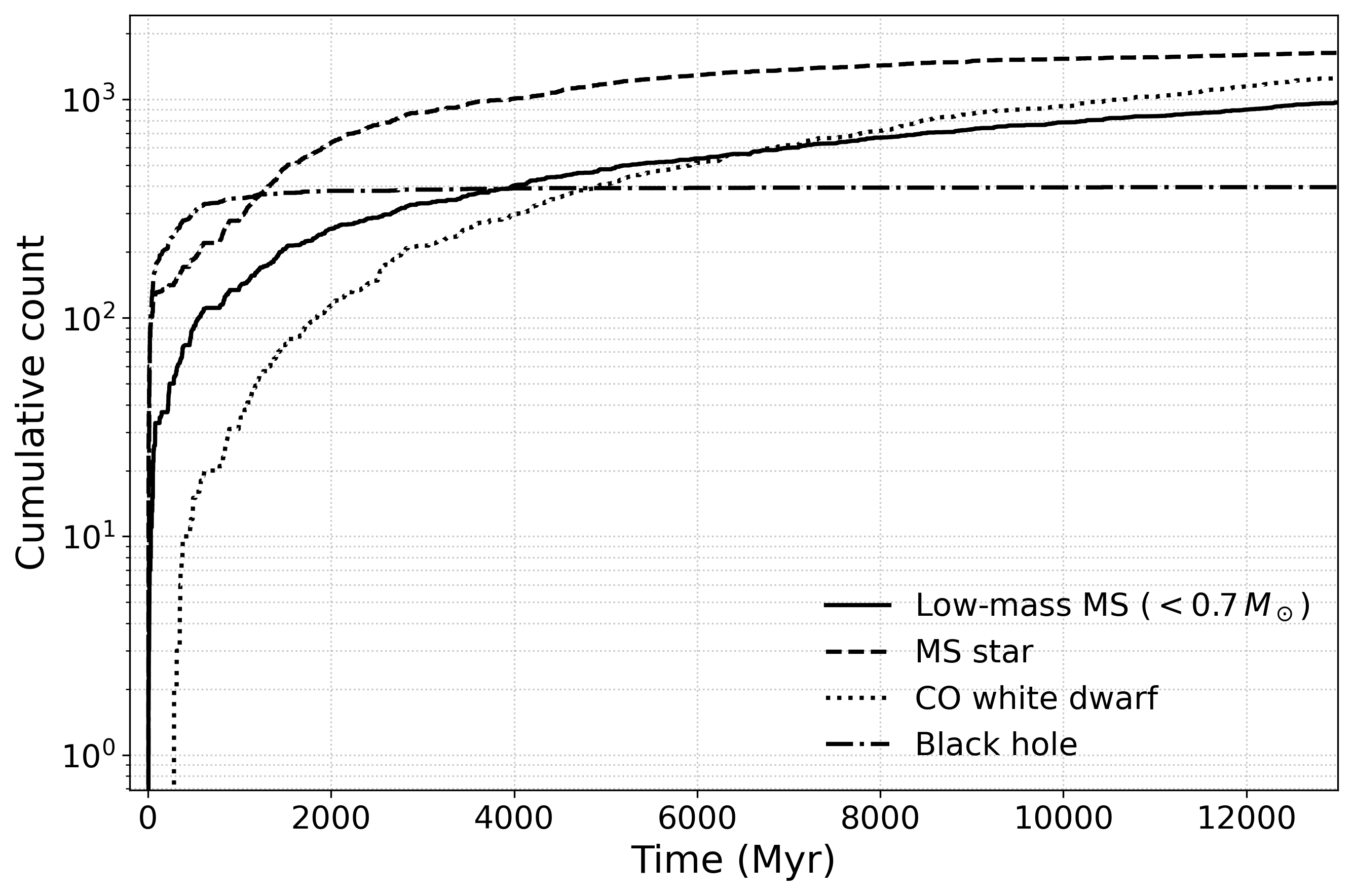}}
        \caption{Cumulative number of collisions/mergers of the central IMBH with different stellar types over 15 Gyr in model rh0.5--W08: low-mass MS stars ($\rm <0.7\,M_\odot$), higher-mass MS stars ($\rm >0.7\,M_\odot$), carbon–oxygen white dwarfs, and sBHs.}

        \label{fig:imbh_mergers}
    \end{subfigure}
    \caption{Formation and growth of IMBHs in MOCCA models. Left: rapid seeding via VMS–sBH collisions and early sBH mergers. Right: long-term growth from MS stars, WDs, and BHs.}
    \label{fig:imbh_mass_evolution}
\end{figure*}

IMBHs form only in the three densest models ($\rm\rho_c\gtrsim10^7\,M_\odot \, \mathrm{pc}^{-3}$; Table~\ref{tab:mocca_models}), underscoring the strong dependence on central density \citep{Sharma2025} and half-mass relaxation time . As shown in Fig.~\ref{fig:imbh_mass_evolution}, a VMS forms within the first few Myr and seeds an IMBH through a VMS–BH collision at $\sim3$ Myr. Over the next 0.5 Gyr, the IMBH grows by a factor of $\sim3$–4, primarily through mergers with sBHs, supplemented by mergers with stars and other compact remnants. IMBH–sBH mergers both drive this growth and rapidly deplete the sBH population, with 524, 390, and 178 such events in models rh0.25--W08, rh0.5--W08, and rh1.0--W08, respectively. These mergers produce ``light IMRIs'' (LIMRIs), most of which have characteristic strains and frequencies within the sensitivity band of next-generation GW detectors such as the Einstein Telescope. At later times, tidal disruption events contribute an additional $\sim10^3$–$\rm 10^4\,M_\odot$ to the IMBH, with massive MS stars dominating early and white dwarfs at later epochs (Fig.~\ref{fig:imbh_mergers}).

\section{Conclusions and Outlook}

Our $N$-body and MOCCA studies show that collisional runaway in extremely dense clusters can produce IMBHs within a few Myr, though growth histories depend on collision prescriptions, rejuvenation, and the internal structure of the VMS. IMBHs form only in the most compact models ($\rm\rho_c \gtrsim 5\times10^6\,M_\odot\,\mathrm{pc}^{-3}$ and $T_{rh} \lesssim 500\,\mathrm{Myr}$), underscoring their sensitivity to cluster density and relaxation time. While both methods agree on the overall pathway, they differ in the number of collisions and the timing of collapse, reflecting their respective approximations. These results assume monolithic cluster formation, which may not be universal. Key uncertainties remain, including the structure and evolution of the VMS and the use of simplified stellar evolution prescriptions. Continued cross-validation of $N$-body and Monte Carlo approaches \citep{vergara2025b}, coupled with improved VMS physics, will refine predictions and guide searches for IMBH signatures in tidal disruptions and GW sources such as LIMRIs.

\medskip
\noindent\textbf{Acknowledgements.} We thank M. Giersz, R. Spurzem, F. Flammini Dotti, A. Kamlah, D. Schleicher, M. Arca Sedda, A. Hypki, J. Hurley, P. Berczik, A. Escala, N. Hoyer, N. Neumayer, X. Pang, A. Tanikawa, R. Cen, T. Naab, G. Wiktorowicz, L. Hellström, and J. Szyndler. AA and SA acknowledge support from project No. 2021/43/P/ST9/03167 co-funded by the Polish National Science Center (NCN) and the EU Horizon 2020 Marie Skłodowska-Curie grant agreement No. 945339. For the purpose of Open Access, the authors have applied for a CC-BY public copyright license to any Author Accepted Manuscript (AAM) version arising from this submission. MCV acknowledges funding through ANID (Doctorado acuerdo bilateral DAAD/62210038) and DAAD (funding program number 57600326). MCV acknowledges the International Max Planck Research School for Astronomy and Cosmic Physics at the University of Heidelberg (IMPRS-HD).

\end{document}